\documentclass[fleqn,usenatbib]{mnras}

\usepackage{mathptmx}

\usepackage[T1]{fontenc}
\usepackage{ae,aecompl}

\usepackage{graphicx}	
\usepackage{amsmath}	
\usepackage{amssymb}	
\usepackage{lscape} 
\usepackage{url} 

\newcommand{\Msun}{$\mathrm{M_{\sun}}$}

\newcommand{\kms}{km\,s$^{-1}$}

\newcommand{\OII}{[\ion{O}{ii}]}
\newcommand{\OIII}{[\ion{O}{iii}]}
\newcommand{\Halpha}{H$\alpha$}
\newcommand{\Hbeta}{H$\beta$}
\newcommand{\NII}{[\ion{N}{ii}]}

\newcommand{\NeIII}{[\ion{Ne}{iii}]}
\newcommand{\NeV}{[\ion{Ne}{V}]}

\title[Star formation or EELR?]{[O\,{\sc{ii}}]\ as a proxy for star formation in AGN host
  galaxies: beware of extended emission line regions}

\author[N. Maddox]{
Natasha Maddox $^{1}$\thanks{E-mail: maddox@astron.nl}
\\
$^{1}$ASTRON, the Netherlands Institute for Radio Astronomy, Postbus 2,
7990 AA, Dwingeloo, The Netherlands\\
}

\date{Accepted XXX. Received YYY; in original form ZZZ}

\pubyear{2018}

\begin{document}
\label{firstpage}
\pagerange{\pageref{firstpage}--\pageref{lastpage}}
\maketitle

\begin{abstract}

The \OII\ 3726$+$3728\AA\ emission line doublet is often used to
estimate star formation rates within the host galaxies
of active galactic nuclei (AGN), as it is known to be strongly excited
by star formation, but is only weakly excited in the broad and narrow
line regions of AGN. However, within AGN host galaxies, \OII\ can also
be excited in low-density gas located at appreciable 
distances from the nucleus, but still ionized by the 
AGN. These AGN extended emission line regions (EELRs) can contribute
significant flux to integrated spectra, even in the presence of luminous AGN.
Here, we identify EELRs by the presence of the \NeV\ 3426\AA\ emission
line, which, like \OII, is not strongly excited in the inner regions
of AGN, but is a prominent emission line in the lower density
EELRs. Critically, unlike \OII, \NeV\ is not excited by star
formation. Therefore, when strong \NeV\ is present in an AGN spectrum, 
the flux from the EELR is not negligible, implying the \OII\ flux 
is contaminated by emission from the EELR, and is not a good measure
of star formation. After removing objects with EELRs identified by
\NeV, the \OII\ flux in the host galaxies of radio-loud AGN
is found to be higher than that within radio-quiet AGN, which could
either indicate higher star-formation rates, or the presence of
moderate-velocity shocks. Being mindful of EELRs for upcoming
large-area spectroscopic surveys, particularly those tied to radio
continuum surveys, will be important for determining star formation
rates in AGN host galaxies.
\end{abstract}



\section{Introduction}\label{sec:Introduction}

The current star formation rate (SFR) is an important observational
metric of a galaxy. It tells us the evolutionary state by indicating
whether a galaxy is actively building its stellar mass, or if it is
passively ageing. Star formation (SF) can also be triggered by, for
example, mergers and/or accretion of gas, thus providing information
about a galaxy's environment. The SF history is a crucial underlying factor in
establishing the diversity of galaxies we observe throughout the universe.

There are many observables at a variety of wavelengths that serve as
calibrated proxies for the current SF in galaxies. These include
optical emission lines such as \Halpha, and continuum flux at
ultra-violet, far-infrared and radio wavelengths. A comprehensive
comparison of various SF tracers can be found in \citet{Hopkins2003}.

The relationship between the flux in the \OII\ 3726+3728\AA\ emission
line doublet (hereinafter referred to as \OII)  and the
SFR has been calibrated by a number of authors (e.g. \citealt{Kennicutt1998},
\citealt{Hopkins2003}, and \citealt{Kewley2004}). Although it suffers
from dust extinction due to the relatively short emission wavelength, \OII\ is very
useful for estimating SFRs, as it is a strong, easily identified feature and can be seen
in moderate resolution optical spectra out to high ($z<1.5$) redshifts. 

Complicating the measurement of SF in some galaxies is the presence of
active galactic nuclei (AGN), as the flux from the AGN outshines the stellar
component, at some wavelengths by orders of magnitude. In order to
measure SF in AGN host galaxies, we require a quantity whose
production is dominated by SF rather than AGN. As shown in the quasar
composite of \citet{VandenBerk2001}, \OII\ is only weakly excited in the
narrow line region (NLR) of AGN, and not at all in the broad line
region (BLR). However, it is seen in emission from even moderate
levels of star formation. It has therefore been used to estimate the
SFR in quasar host galaxies (e.g. \citealt{Ho2005},
  \citealt{Kim2006}, \citealt{Kalfountzou2012}, 
\citealt{Matsuoka2015}, and \citealt{Vergani2017}).

\subsection{Extended Emission Line Regions}

AGN are sometimes seen to ionize gas in the host galaxy beyond the BLR
and NLR. While the NLR is typically confined to within $<$1 kiloparsec
(kpc) of the nucleus, extended emission line regions (EELRs)
\footnote{We use the term EELR to describe extended emission beyond the NLR
which, based on emission line diagnostics, is shown to be gas ionized
by the AGN. We do not use the sometimes quoted alternative term
`extended narrow line region' (ENLR), to reinforce the fact that the
gas is distinct from the NLR.} can
extend throughout the entire host galaxy, spanning tens of kpcs in
some cases (e.g. \citealt{Fosbury1982}, \citealt{Spinrad1984},
\citealt{Stockton1987}, \citealt{Fu2009a}, \citealt{Villar2011},
\citealt{Husemann2013}, \citealt{Liu2013a}, \citealt{Liu2013b},
\citealt{Liu2014}, \citealt{Harrison2014},
\citealt{Husemann2014}). EELRs are spatially and kinematically 
distinct from the classic AGN NLRs. The gas within an EELR can show
velocity differences of several hundreds, to $>$1000 \kms\ with
respect to the systemic velocity of the galaxy, but also very low
velocity dispersion (\citealt{Fu2009a}, \citealt{Husemann2013}).

Early studies employing long-slit spectroscopy focused on the EELRs
within the host galaxies of bright radio sources
(\citealt{Spinrad1984}, \citealt{Unger1987}), investigating possible 
triggering mechanisms for the nuclear activity, including mergers with
gas-rich galaxies. However, EELRs are also now known to be observed in
AGN hosts with no appreciable nuclear radio emission and no indications
of recent merger activity (e.g. \citealt{Villar2011}, \citealt{Husemann2013}). 

Integral field unit (IFU) spectrographs with wide spectral range and
substantial (tens of arcseconds) fields-of-view have been very useful for studying the
kinematics and spatial extent of EELRs. \citet{Fu2009a} observed eight
galaxies known to have extended emission, covering a wide wavelength
range, to investigate EELR clouds on small spatial scales. The ratios
of their detected emission lines all indicate AGN as the source of the
ionizing flux, rather than star formation. Similarly,
\citet{Husemann2013}, \citet{Liu2013a}, \citet{Harrison2014} and
\citet{Husemann2014} also use IFU observations to isolate regions
throughout AGN host galaxies where the gas has been ionized by the AGN.

Many EELR studies focus on optical emission lines such as \OIII\ 5008\AA\ and
\Halpha. However, for studies with wider wavelength coverage, or
observations of objects at higher redshifts, shorter wavelength lines
are also visible throughout the EELR, including strong \OII.
The lower density of the interstellar
gas, of a few hundred cm$^{-3}$ (\citealt{Husemann2014}), with respect
to the order of magnitude higher density of the AGN NLR, enables \OII\ emission,
which has a low critical density, to arise within the EELR (\citealt{Villar2011}). 
\citet{Fu2009b} observed a subset of their EELR galaxies in the
infrared (IR), but found no signatures of SF in the hosts, thus
confirming the \OII\ flux detected in the IFU spectra should be
attributed to the EELR, rather than SF. 

As both SF and AGN EELRs independently result in strongly excited
\OII\ throughout the full extent of galaxies via unrelated mechanisms,
we require some diagnostic to distinguish 
between the different ionizing sources. While emission line ratios can
be broadly used to determine the source of the ionizing radiation (i.e. AGN or
SF), the most commonly used spectral features, including \Halpha, are
shifted out of the optical spectroscopic range by moderate, $z\sim
0.4$, redshifts. 

Fortunately, there are specific emission lines which are weak in both
the high-density medium surrounding AGN as well as star-forming 
regions. One such emission line is \NeV\ 3426\AA, which, due to its
high ionization potential (97 eV, compared to 13.6 eV for \OII,
\citealt{Lide2005}) is not excited by star formation, but as with
\OII, is very weak in AGN spectra \citep{VandenBerk2001}.  
Indeed, for observations of EELRs with spectral coverage of these
short wavelengths, \NeV\ is often seen as well
(e.g. \citealt{Spinrad1984}, \citealt{Fu2009a}). 

Here we investigate the subset of AGN which, in addition to
emission lines common in either SF or AGN spectra, also show strong
\NeV\ in their spectra, and argue that it indicates the presence of an
EELR, or in some cases, extreme shocks. Since EELRs also strongly
excite \OII, the \OII\ in these objects should not be used as a 
proxy for SF, since the flux in this emission line is contaminated.
In Section~\ref{sec:ratios} we investigate the diagnostic power of
emission line ratios in determining the properties of the ionizing
radiation, and where objects with detected \NeV\ emission lie in this
parameter space. Section~\ref{sec:case} is a case study of star
formation in SDSS quasar host galaxies. A discussion and conclusions
are presented in Section~\ref{sec:discussion}. Concordance
cosmology with $H_{0} = 70$ km s$^{-1}$ Mpc$^{-1}$ (thus $h\equiv
H_{0}$/[100 km s$^{-1}$ Mpc$^{-1}$]$=0.7$), $\Omega_{m} = 0.3$,
$\Omega_{\Lambda} = 0.7$ is assumed throughout. We use laboratory
wavelengths for emission lines, taken from Table 2 in \citet{VandenBerk2001}.

\section{Emission line diagnostics}\label{sec:ratios}

\begin{figure}
\includegraphics[width=\columnwidth]{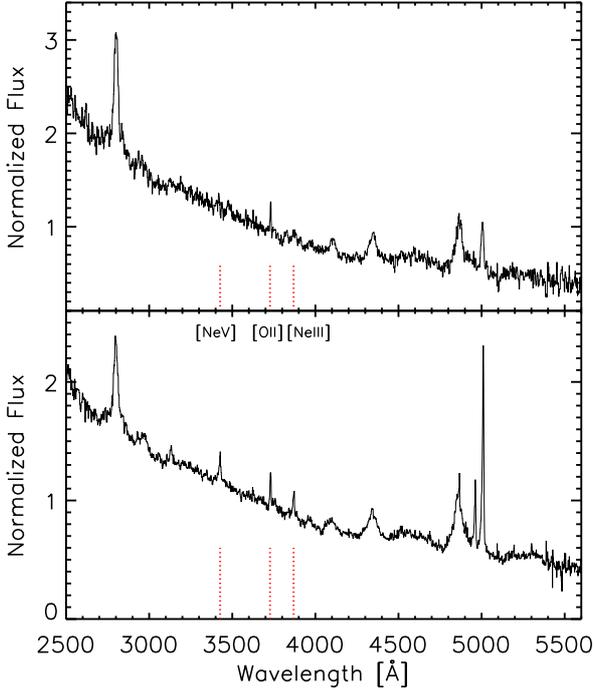}
\caption{Two SDSS spectra of AGN at $z\sim 0.63$, shifted to the
  rest-frame, showing the range of \OII\ and \NeV\ properties under investigation. The top
  spectrum has a prominent \OII\ emission line but no \NeV, while the
  bottom spectrum has strong \OII\ emission with similarly strong
  \NeV. \NeIII\ is also seen redward of \OII. The locations of \NeV,
  \OII\ and \NeIII\ are marked along the x-axis with red dotted lines.}
\label{fig:2spec}
\end{figure}

The emission lines we are interested in are the \OII\ doublet, and
\NeV\ 3426\AA\ (referred to as \NeV). We also investigated
\NeIII\ 3869\AA\ (referred to as \NeIII) but found it to have less
diagnostic power than \NeV. 
Fig.~\ref{fig:2spec} shows the variation in emission strengths
of these species in two illustrative spectra of AGN selected by the
Sloan Digital Sky Survey (SDSS, \citealt{York2000}). The top spectrum
is dominated by flux from the AGN, and the unusually strong \OII\
indicates additional flux from SF. The bottom spectrum shows \OII, but also
has equally prominent \NeV\ shortward of \OII\ (and \NeIII\ longward),
which are not strongly excited within the high density broad or narrow line regions
of AGN, or by SF activity.

\subsection{BPT classification}

\begin{figure}
\includegraphics[width=\columnwidth]{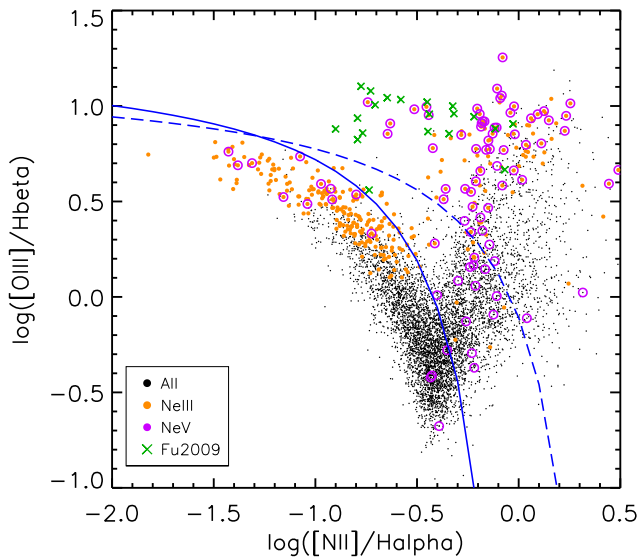}
\caption{BPT emission line diagnostic diagram for SDSS objects at
  $0.2<z<0.25$ (small black points). The blue solid line is the
  empirical separation between AGN and SF ionizing radiation, as determined
  by \citet{Kauffmann2003}, while the blue dashed line is the
  theoretical maximum for SF from \citet{Kewley2001}. Points above and 
  to the right of the dashed demarcation line are dominated by AGN-type
  radiation, while objects between the dashed and solid lines have a
  mix of ionizing sources. Objects with \NeIII\ or \NeV\ additionally detected in
  their spectra are marked with orange points and purple circles,
  respectively. The \NeV\ objects lie almost exclusively in the AGN
  region. Objects with \OII\ in their spectra span the full parameter space.
  The green crosses are the EELR galaxies from \citet{Fu2009a}.}
\label{fig:BPT1}
\end{figure}

Ratios of flux from various emission lines can be used to understand
the nature of the ionizing radiation exciting the emission
(e.g. \citealt{Baldwin1981}, \citealt{Robinson1987}). With specific
combinations of spectral features, clear separation between blackbody
radiation from stars and radiation with a power-law spectrum is
found. The most common of these emission line ratio plots is shown in
Fig.~\ref{fig:BPT1}, often referred to as the `BPT diagram', used here
for illustration, where the ratio of \OIII\ 5008\AA\ to \Hbeta\
4862\AA\ is plotted against \NII\ 6585\AA\ to \Halpha\
6564\AA (hereinafter referred to as \OIII, \Hbeta, \NII\ and
\Halpha). These two pairs of emission lines are close in 
wavelength, therefore the ratios are insensitive to dust. 

For Fig.~\ref{fig:BPT1}, we extracted objects from the SDSS database
table \texttt{emissionLinesPort}, which has measured a large number of
spectral features. We use the following lines for investigation:
\NeV\_3425, \OII\_3726+3728, \NeIII\_3868, \Hbeta\_4861,
\OIII\_5006, \Halpha\_6562 and \NII\_6583. We extract only objects within the redshift
range of $0.2<z<0.25$ to ensure spectral coverage ranging from \NeV\ 
through \NII. We imposed no other restrictions on spectral 
classification or magnitude, but every object is morphologically
extended and bright enough to have been selected for spectroscopic follow-up. The
resulting population is a mix of galaxies with a range of SFR and AGN of low enough
luminosity that the host galaxy is still visible.

Two sequences are immediately seen, separated by the demarcation
between AGN and SF as determined by \citet{Kauffmann2003}, indicated
by the solid blue line in Fig.~\ref{fig:BPT1}. The less strict
demarcation from \citet{Kewley2001} is also shown as a dashed blue
line. Objects below and to the left of the solid line are excited by radiation from 
SF. The sequence within the SF locus toward lower values of
\NII/\Halpha\ follows decreasing metallicity (\citealt{Kewley2001}). Objects lying above and
to the right of the dashed line are generally considered to be ionized
by radiation consistent with an 
AGN, with objects that fall between the two lines classified as
`transition objects', subjected to a mix of ionization sources.

To illustrate the diagnostic power of \NeV, and to a lesser extent,
\NeIII, to identify EELRs, we highlight the objects with each of these
emission lines detected in their spectra. Purple circles mark objects
for which \NeV\ is measured at $\ge$5$\sigma$. They are almost exclusively
within the AGN region, indicating SF is not responsible for exciting this
species. Indeed, other works have used the \NeV\ line alone to select
samples of AGN from large spectroscopic samples
(e.g. \citealt{Mignoli2013}, \citealt{Vergani2017}), arguing that it
unambiguously indicates the presence of hard ionizing radiation.

There is a small cluster of points at the low-metallicity end of the star
  formation sequence with detected \NeV. The low-metallicity AGN models of
  \citet{Feltre2016} populate this region of parameter space, for a
range of AGN spectral index and ionization parameters.
Orange points are objects with $\ge$5$\sigma$ \NeIII\ emission
lines. They are primarily confined to the low-metallicity branch of the SF sequence,
but also extend into the AGN cloud. Objects with detected \OII\ (not
highlighted on this figure) span the 
full parameter space, as both SF and AGN EELRs excite this species.

Also marked on Fig.~\ref{fig:BPT1} as green crosses are points from the IFU
observations of EELR clouds from \citet{Fu2009a}. They all lie within
the AGN cloud, and extend toward lower metallicity. Eight of the 20
clouds have measurable \NeV\ emission, but four of the remaining 12
objects do not have spectral coverage shortward of \Hbeta, so at least
40~per~cent of the \citet{Fu2009a} EELR clouds show clear \NeV.

We conclude that the presence of \NeV\ is a nearly unambiguous
indication of an EELR, in agreement with previous work based on
\NeV-selected AGN, whereas \NeIII\ is less reliable, appearing
predominantly in objects with BPT emission line ratios indicative of
low-metallicity SF. However, the lack of detected \NeV\ emission in
$\sim$ half of the \citet{Fu2009a} EELR clouds shows that not every
EELR shows \NeV\ at the same level as \OII.

\subsection{Ionization from shocks}\label{subsec:shocks}

Shocks within the interstellar medium are also capable of exciting
a rich spectrum of emission lines. To investigate whether the \OII\
and \NeV\ we see in the spectra are due to shocks rather than EELRs, 
we employ the MAPPINGS~III shock and photoionization modeling
code and results from \citet{Allen2008}. A comprehensive library of models with
varying magnetic field strength, density, metallicity and shock velocity were
constructed. The expected fluxes of various species from not 
only the radiative shock, but also the photoionized precursor medium,
along with a combination of the two (shock $+$ precursor, $s+p$), 
are tabulated. From the models it is clear that a wide variety of shock conditions
are capable of exciting \OII, resulting in yet another process that can contribute to
this emission line independent of SF. 

We focus on combinations of emission line ratios, including \NeIII\
and \NeV, to discriminate between shock and EELR ionization. With
respect to the BPT diagram, none of the precursor-only models are
found within the AGN cloud for any combination of parameters, so they
can be excluded. The models including only the shock does produce
AGN-like BPT emission-line ratios, but no appreciable \NeV\
emission. Therefore, only the $s+p$ models are considered further. These
are most relevant for observations where the distinct regions are
unresolved, as is the case for fibre-based spectroscopy. 

All $s+p$ models from the \citet{Allen2008} models except those with
SMC-like or LMC-like abundances approximately lie in the AGN region of
the BPT diagram. These models also produce \NeV\ at high ($>$600\kms)
shock velocities. However, only at the highest modeled velocities,
$\sim$900\kms, are the \NeIII/\NeV\ ratios small, i.e. $\lesssim$5,
typical of the ratios measured in the spectra described in
Section~\ref{sec:case}. Therefore, we find that only the $s+p$ 
models at the highest velocities are capable of producing emission
line ratios consistent with the AGN region of the BPT diagram while
also producing \NeV\ and \NeIII\ in the correct quantities. The region
of the BPT diagram covered by these very specific shocks is restricted
to the top right corner of Fig.~\ref{fig:BPT1}, not spanning the full
AGN cloud as shown by the \NeV-detected objects. We therefore conclude
that while very high velocity shocks are capable of exciting \NeV, it
is not the dominant mechanism.

The high shock velocities, $\sim$900\kms, required to produce \NeV\ 
imply they are driven by an AGN, and are possibly, but not
exclusively, related to radio jets. Therefore, both EELRs and high-velocity
shocks result from AGN activity, albeit via different mechanisms. The
essential point is that whether via direct photoionization from the
AGN or ionization from AGN-driven shocks, the production of \NeV\ is
unambiguously related to AGN activity acting on gas beyond the BLR and
NLR. 


\section{Case Study: Star formation in AGN hosts}\label{sec:case}

Here we revisit the appropriateness of using \OII\ to derive the SFR in AGN host
galaxies, and investigate the effect of excluding objects which show
\NeV\ emission due to EELRs. We restrict our analysis to $0.1<z<1.3$,
with the lower and upper redshift limits set by requiring coverage of
\NeV\ and \OII. We use the SDSS Data Release 7 (DR7,
\citealt{Abazajian2009}) quasar catalogue (\citealt{Schneider2010}),
which has high completeness for 
quasars at these redshifts. Subsequent data releases focused
primarily on higher redshift targets, so DR7 is currently the best
dataset for this investigation.

No database tables within SDSS measure our required emission lines
for quasars over our full redshift range, so we 
measure the emission lines ourselves. We have downloaded all the
spectra for quasars within the DR7 quasar catalogue at $0.1<z<1.3$. We
measure \NeV\ and \OII\ directly from the spectra, by fitting
a Gaussian profile in a spectral window 150\AA\ wide centred on the expected
line wavelengths. Our measured fluxes and continuum levels for \OII\
match very well for objects with measurements in the
\texttt{galSpecLine} table, which are confined to $z<0.7$. Note that
we fit the combined \OII\ doublet with a single Gaussian instead of
separating it into two components. 
Based on visual examination of the resulting fits, and comparison with
the SDSS catalogue values, an emission line is considered to be
detected if the line is successfully fit automatically, and the
equivalent width (EW) is confident if the flux in the line is measured
to be $\ge 10^{-16}$ erg s$^{-1}$ 
cm$^{-2}$. This flux limit does not result in an effective lower limit
on the computed EW, provided the continuum level is sufficiently high.

\subsection{[O\,{\sc{ii}}] Equivalent widths}

\begin{figure}
\includegraphics[width=\columnwidth]{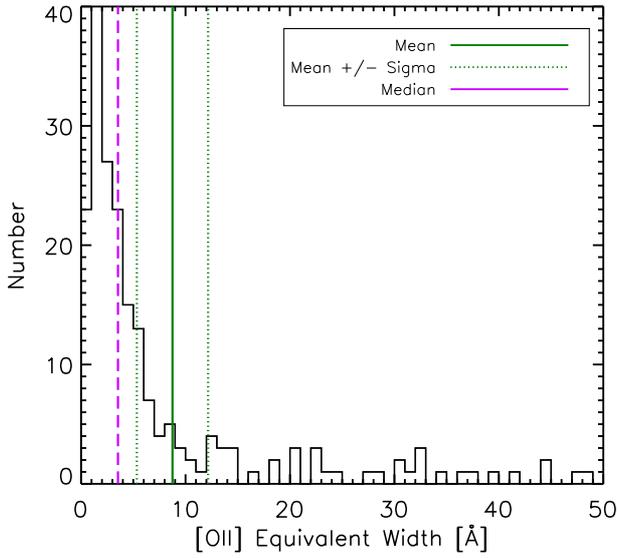}
\caption{Distribution of equivalent widths of the \OII\ emission line for quasars in
  a single redshift bin. The green solid line is the mean of the
  distribution, while the purple dashed line is the median. The green
  dotted lines are the mean $\pm$ the sigma of the distribution.}
\label{fig:EWmeasures}
\end{figure}

Fig.~\ref{fig:EWmeasures} shows the distribution of \OII\ EW 
measured for quasars in the $0.5<z<0.6$ redshift bin,
converted to the rest-frame. The
green vertical line marks the mean of the distribution, which is
clearly affected by the small number of objects with high EWs. The
green dotted lines show the standard deviation from the mean, which is
not strictly meaningful for this type of distribution, but gives an
idea of the spread of values. The purple dashed line shows the median
of the distribution, which is more robust against the tail toward high
EW values. Since it is the high \OII\ EW objects that are most often
associated with EELRs, we use the mean and standard deviation to parametrize the
EW distributions, with the understanding that the values are
illustrative rather than statistically robust.

After measuring the \OII\ and \NeV\ emission lines for every spectrum,
we divide the quasars into redshift bins of width $\Delta z=0.1$, 
and then divide the quasars in each redshift bin into two groups,
based on their radio properties. We use the definition from 
\citet{Ivezic2002} for radio-loudness, R$_i$, using the Faint Images
of the Radio Sky at Twenty-centimeters survey (FIRST, \citealt{Becker1995}) peak flux
density provided in the DR7 quasar catalogue. Objects with R$_i > 1$
are classified as radio-loud (RL), while objects with R$_i < 1$ are
radio quiet (RQ).

In order to compare the properties of RL and RQ quasars, we create
samples matched in redshift and observed $i$-band magnitude to avoid
any biases due to luminosity. For each RL quasar in a redshift bin, we
choose the RQ quasar nearest in redshift and magnitude, with matching
tolerance of 0.05 in $z-i$-band magnitude space. We do this both for
the full populations, as well as the sub-populations where
the \OII\ emission line is detected but the \NeV\ emission line is not
detected in the individual spectra. From Fig.~\ref{fig:2spec}, the
full population would include both objects in the figure, whereas
the sub-population would only include the object shown in the top panel.

\begin{figure}
\includegraphics[width=\columnwidth]{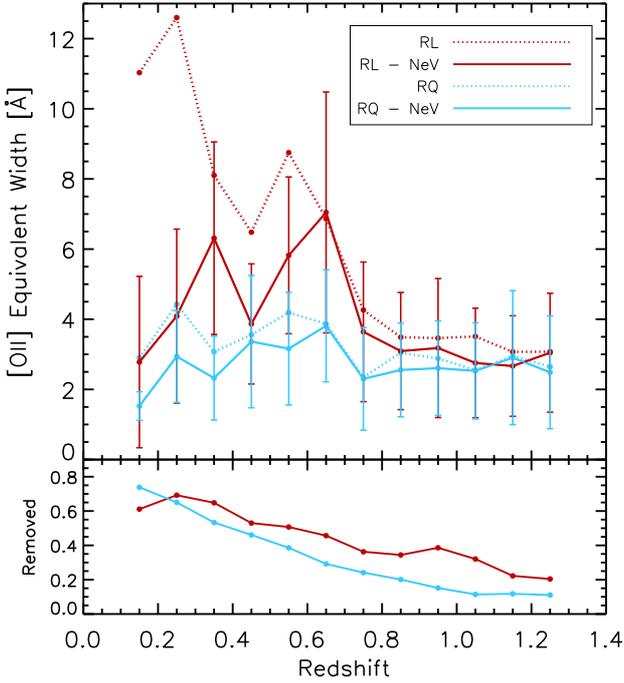}
\caption{Equivalent width of the \OII\ emission line as a function of
  redshift, for RL (red) and RQ (blue) quasars. The dotted lines are
  average EW values for each redshift bin, while the solid lines are
  average EW values for each redshift bin excluding objects with \NeV\
also detected in the spectrum. The lower panel shows the fraction of
objects excluded from each bin for having \NeV\ detected in individual spectra.}
\label{fig:matchedEW}
\end{figure}

Fig.~\ref{fig:matchedEW} shows the mean of the derived \OII\ EW
for the matched samples as a function of redshift, in redshift bins $\Delta z=0.1$, separated
into RL (red) and RQ (blue). When we include all RL and RQ quasars,
shown as dotted lines, the RL 
quasars have much larger \OII\ EW than the RQ quasars, to
$z=0.9$. When we exclude objects with detected \NeV, which imply the
\OII\ fluxes are contaminated by emission from the EELR, the average
EW for RL and RQ are both reduced, and shown as the solid red and blue
lines, respectively. The two lines are also much closer, indicating a
smaller difference between the \OII\ EW in the RL and RQ
populations. Above $z>0.9$, the RL and RQ subsets have
indistinguishable mean \OII\ EW, irrespective of the \NeV\
correction. The `error bars' are the illustrative standard deviations, as shown
in Fig.~\ref{fig:EWmeasures}. A two-sided Kolmogorov-Smirnov (KS) test
on the resulting mean EW values for the RL and RQ populations for each
redshift bin, shows that the two full populations stop being
  significantly different at $z=0.7$, but the uncertainties on the
  measurements are such that the samples of objects with \NeV\ removed
are not formally different at any redshift.

The decrease in \OII\ EW in the $0.4<z<0.5$ redshift bin for the RL
quasars may be a result of the \Hbeta\ emission line entering the SDSS
$i$-band. The SDSS quasar catalogue is flux-limited in the
$i$-band, and the prominent \Hbeta\ emission line entering the band
artificially boosts the apparent magnitude of the objects, resulting
in intrinsically less luminous quasars satisfying the flux
limit. The fact that the RL quasars are most affected also indicates
that the radio properties of the quasar targets also affected their
selection, as objects with detections in FIRST are preferentially
targeted by the SDSS quasar selection algorithm. 

The bottom panel of Fig.~\ref{fig:matchedEW} shows the fraction of
objects with measurable \NeV\ in their spectra, and are thus removed
from the sample. At $z<0.3$, we find the fraction of spectra showing \NeV\ in their
spectra is similar for the RL and RQ sub-populations, but at higher
redshift, the fraction is higher in RL hosts by 5--15~per~cent. 
The fraction of objects with measurable \NeV\
steadily decreases with increasing redshift, reflecting the fact
that at higher redshift, the intrinsically more luminous AGN
increasingly dominate the flux in the spectra, with the average
luminosity of the objects in the $z=1$ bin an order of magnitude
more luminous than in the $z=0.1$ bin. At $z\sim 1$, the
\NeV\ fractions flatten at around 10~per~cent for RQ 
and 20~per~cent for RL. This shows that EELRs (or in addition, strong shocks) are
more common within RL quasars, but also visible in a fraction of RQ quasars.

We keep our results as a function of redshift because of the known strong
evolution in SFR from $0<z<1$, and the increase in intrinsic
luminosity of the quasars within the DR7 quasar catalogue. If we
average over the entire $0.1<z<1.3$ redshift range, we find uncorrected
average \OII\ EW = 5.8 and 3.4 for RL and RQ quasars,
respectively. Removing the \NeV-detected objects give corrected mean
\OII\ EW values of 4.0 and 2.8, for the RL and RQ quasars, respectively.
These are smaller than the values quoted in \citet{Kalfountzou2012},
who find average \OII\ EW for RL and RQ quasars 7.80 and 4.77,
respectively. However, note that \citet{Kalfountzou2012} impose a lower limit of
EW$=3$ to be included in their sample, thus biasing the result to
larger values.

\subsection{Stacked spectra}

\begin{figure}
\includegraphics[width=\columnwidth]{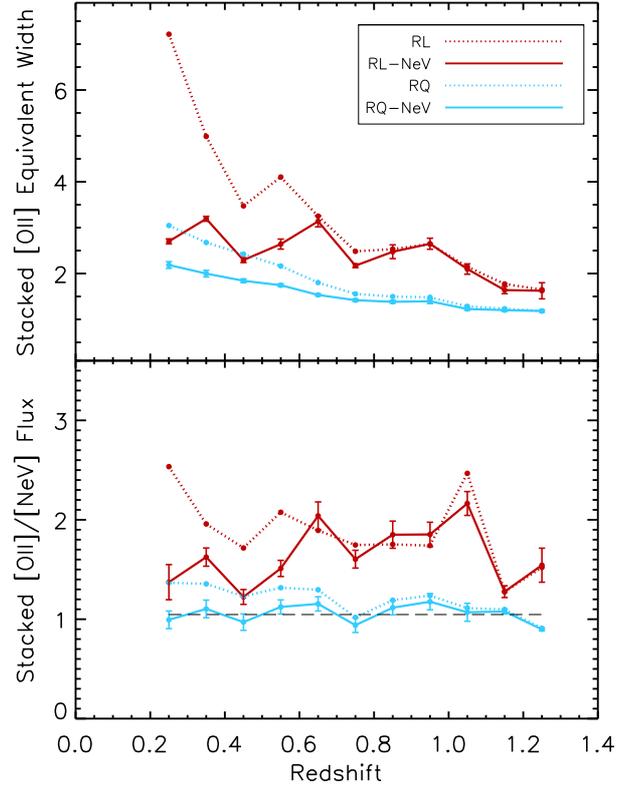}
\caption{(Top) Equivalent width of the \OII\ emission line as a function of
  redshift, for RL (red) and RQ (blue) quasars. The dotted lines are
  EW values for the stacked spectrum for each redshift bin including
  all the spectra, while the solid lines are EW values for the stacked
  spectrum for each redshift bin excluding objects with \NeV\ also
  detected in the individual spectra. (Bottom) Ratio of \OII\ to \NeV\
  fluxes, from the stacked spectra. Also shown as the horizontal black dashed line is the flux
  ratio of the two emission lines from the SDSS quasar composite
  spectrum.}
\label{fig:stackedEW}
\end{figure}

We stack the SDSS spectra in bins of redshift of width $\Delta z=0.1$,
using the noise-weighted 
mean, in order to investigate the mean spectral properties of objects
in each redshift bin, along with weak spectral features not identified
in individual spectra. We again separate the objects into RL and RQ
subsamples. In addition, we create two subsamples for each redshift
bin; one containing all objects in the redshift bin, and one
containing only objects where \NeV\ is not detected in individual
spectra. This results in four stacked spectra per redshift bin. 
We then measure the \OII\ EW in the stacked spectra. The
results are shown in the top panel of
Fig.~\ref{fig:stackedEW}. Uncertainties are derived from the
uncertainty in placing the continuum level, along with the uncertainty
in the fit of the emission line, and are small due to the high
signal-to-noise ratio (SNR) of the stacked spectra. The
stacked \OII\ EW measurements show qualitatively similar behaviour
as for the individual spectra in Fig.~\ref{fig:matchedEW}. The high
SNR of the stacked spectra enable secure
measures of smaller EW values, reaching EW$\sim$1 at the highest
redshifts probed.

The SDSS composite quasar spectrum from \citet{VandenBerk2001} shows
that the \OII\ and \NeV\ emission lines are both intrinsically weak,
and at similar flux levels, with the ratio between the two of
\OII/\NeV=1.05. We investigate the ratio of \OII/\NeV\ in the stacked
spectra, as shown in the bottom panel of Fig.~\ref{fig:stackedEW}. Even
after excluding quasars with \NeV\ detected in individual spectra,
\NeV\ is detected in the higher SNR stacked spectra.  The \OII/\NeV\
ratio for RL quasars is $\sim$2, much larger than 1.05, indicating
stronger \OII\ than would be 
expected if it was produced by the AGN alone. This corroborates the
weak excess seen in Fig.~\ref{fig:matchedEW}, that RL quasar hosts
have elevated \OII\ with respect to RQ hosts. The RQ hosts have emission
line ratios consistent with the ratio from the SDSS quasar composite,
indicating the emission lines measured in the stacked spectra are
dominated by the AGN NLR.

\subsection{Star formation in AGN hosts}\label{subsec:sfstack}

\begin{figure}
\includegraphics[width=\columnwidth]{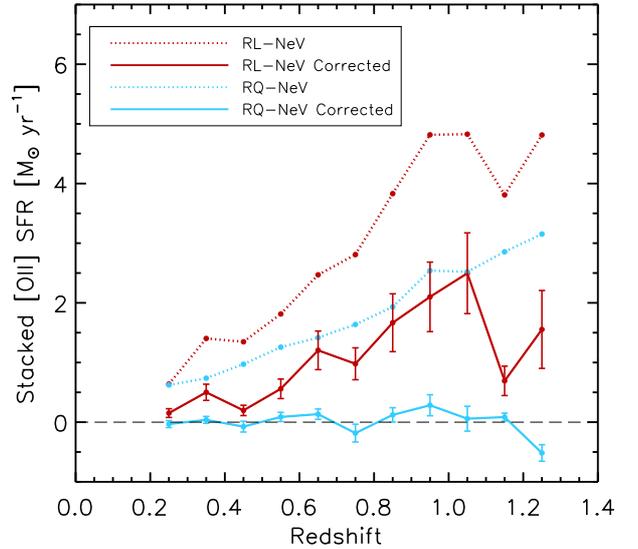}
\caption{SFR derived from the \OII\ flux from the stacked spectra. The
dotted lines convert the total measured \OII\ flux from each stacked
spectrum into a SFR, whereas the solid lines have the \OII\ flux
intrinsic to the quasar subtracted off. The corrected SFR for RL quasars is non-zero
at all redshifts, whereas the corrected SFR for the RQ quasars is
consistent with zero.}
\label{fig:stackedSF}
\end{figure}

The \OII\ fluxes from the stacked spectra can be converted into a SFR
using the conversion from \citet{Kewley2004}. We use only the stacked
spectra that have had objects with \NeV\ visible in individual spectra
removed, corresponding to the solid lines in
  Fig.~\ref{fig:stackedEW}, thus the flux should be free from
  contamination from EELRs. First, we measure the \OII\ flux and
directly convert that to 
a SFR for the RL and RQ subsets, which are shown as red and blue
dotted lines in Fig.~\ref{fig:stackedSF}. The SFR values we derive are
broadly consistent with those from \citet{Matsuoka2015}, who determine
SFR of broad-line quasars using \OII\ measured from high SNR spectra,
with values ranging between 1--10 \Msun\, yr$^{-1}$ over $0<z<1$.
If we use the \OII\ to SFR conversion from \citet{Kennicutt1998}
instead of that from \citet{Kewley2004}, the SFRs derived increase
slightly, but are still consistent with those from \citet{Matsuoka2015}.

However, we know that AGN do indeed produce low levels of \OII\
unrelated to SF, which we must additionally subtract off.
We use the intrinsic \OII/\NeV\ ratio from the SDSS quasar
composite, and the \NeV\ measured in the stacked spectra. Recall that
individual spectra with \NeV\ visible were not included in the stacks,
but due to the resulting high SNR of the stacked spectra, the
weak \NeV\ emission line from the AGN NLR is now visible. We assume
that the \NeV\ flux in the stacked 
spectra is the intrinsic flux from the AGN, and not a result of 
EELRs. From the known intrinsic ratio of \OII/\NeV, we can estimate
the intrinsic strength of the \OII\ line from the AGN NLR and subtract
that from the total flux. The remaining corrected \OII\ flux is then
converted into a SFR as above, and shown on Fig.~\ref{fig:stackedSF}
as the red and blue solid lines. 

The doubly-corrected (once to remove contamination from EELRs and once
to correct for contribution from the AGN NLR) \OII\ flux, when
converted into a SFR, results in a SFR for RQ quasars that is
consistent with zero. This does not mean that there is no star
formation within RQ quasar host galaxies, as we have observed SF
within quasar hosts at low redshifts 
(e.g. \citealt{Kauffmann2003}), intermediate redshifts
(e.g. \citealt{Matsuoka2014}) and $z>1$ (e.g. \citealt{Floyd2013}). 
It only means that when averaged over many objects
within a redshift bin, the SFR is not sufficiently high to produce a
strong enough \OII\ emission line visible beyond the bright quasar
continuum and the flux in the \OII\ line naturally produced within the
AGN NLR.

In contrast, even after removing contamination from EELRs and the
AGN NLR, there is still an excess of \OII\ flux seen in the RL quasar
hosts. This may be due to either star formation in the host galaxies,
or shocks within the host galaxies driven by the radio activity. As we
are unable to differentiate between shocks and SF,
the solid red line in Fig.~\ref{fig:stackedSF} is an upper limit to
the SFR in the host galaxies of RL quasar hosts.

It is important to note that the primary purpose of this exercise is
not to provide a definitive SFR for quasar host galaxies; rather it
is intended to show the magnitude of the difference in computed SFRs
when accounting for flux within the \OII\ emission line
that we know does not come from SF.

\section{Discussion and Conclusions}\label{sec:discussion}

We have shown that strong \NeV\ emission detected in individual spectra is an
indicator of the presence of extended emission excited by
an AGN, either via photoionization, or in some extreme cases, via
high-velocity AGN-driven shocks. This emission line is not strongly
excited within the usual AGN broad or narrow line regions, or by star
formation. Table~\ref{tab:sources} summarizes the relevant sources of ionizing
radiation, and whether they excite \OII\ and \NeV. As \OII\ is also
excited within EELRs, we conclude that for objects with \NeV\
emission, the \OII\ flux is contaminated and cannot be used as a measure of SF.

\begin{table}
\centering
  \caption{Sources of ionizing radiation within AGN host galaxies, and 
  whether \OII\ and \NeV\ are strongly excited by each. $^{\star}$\NeV\ is only
  excited by shocks at high ($>$600\kms) velocities.}
\label{tab:sources}
\begin{tabular}{lcc} \hline
Source & \OII & \NeV \\ \hline
SF &  Strongly & Weakly \\
AGN BLR & $\times$& $\times$ \\
AGN NLR & Weakly & Weakly \\
AGN EELR & Strongly & Strongly \\ 
Shocks & Strongly & Strongly$^{\star}$ \\  \hline
\end{tabular}
\end{table}

The fraction of SDSS DR7 quasars removed due to \NeV\ contamination
ranges from 70~per~cent at low redshift, to 20~per~cent and
10~per~cent for RL and RQ quasars, respectively, at $z=1.3$. After
removing objects with \NeV\ detected in individual spectra, stacking
the remaining spectra reveals a \OII/\NeV\ flux ratio elevated above
that expected from pure AGN emission for the RL quasars, indicating an
excess of \OII\ emission within these objects. The excess of
\OII\ flux in RL quasars can be attributed to star formation, but may
also be partly due to moderate-velocity shocks from the radio jets acting within
the host galaxy, which do not excite \NeV. Therefore, the SFRs shown
in Fig.~\ref{fig:stackedSF} are upper limits.

The fact that the EELR diagnostic line \NeV\ is so close
to \OII\ in wavelength is particularly fortuitous, as they are both
visible over the same redshift range, and when taken as a ratio, it does not suffer
severely from dust reddening, even though they are in the rest-frame
$u$-band. If there is appreciable amounts of dust in the quasar host galaxies,
it will serve to decrease the strengths of both the \OII\ and 
\NeV. Therefore, there may be some objects whose \NeV\ has been
attenuated below the threshold for detection in individual spectra, in
which case these objects were not removed from the stacked samples
used for Fig.~\ref{fig:stackedEW} and Fig.~\ref{fig:stackedSF} and the
\OII\ measurements are still contaminated. However, 
as the SDSS quasars are optically selected, the dust reddening of
individual objects is low (\citealt{Maddox2012}), thus we expect this
effect to be small.

A number of large-area radio continuum surveys are either already
underway, or are about to come online. In the Northern hemisphere the
Low-Frequency Array (LOFAR, \citealt{vanHaarlem2013}) is surveying 2$\pi$
steradians of the northern sky (\citealt{Shimwell2017}). Spectroscopy
for this survey will be supplied by the new spectrograph to be
installed on the William Herschel Telescope (WHT), the WHT Enhanced
Area Velocity Explorer (WEAVE, \citealt{Dalton2014}), providing
redshifts for radio continuum-selected sources
(\citealt{Smith2016}). Covering the Southern hemisphere, the Evolutionary
Map of the Universe (EMU, \citealt{Norris2011}) will be 
undertaken with the Australian SKA Pathfinder (ASKAP, \citealt{Johnston2008}). This
will be coupled with the spectroscopic redshift survey Taipan
(\citealt{daCunha2017}). While the SFR for radio-detected galaxies
can be determined directly from the radio flux, computing SFR from
\Halpha, and to higher redshifts, \OII, will serve as a useful
check. In addition, the radio flux can not be used as a SFR indicator
for radio-loud quasars, leaving \Halpha\ and \OII\ as the only SF
diagnostics available from planned large-area spectroscopic surveys. Other
calibrated SF indicators, such as flux at far-infrared (FIR)
wavelengths, would require additional observations not currently planned. 
The importance of understanding and identifying
the contribution from EELRs to the \OII\ emission line, as measured by
the presence of \NeV, is essential for these projects.


\section*{Acknowledgements}

We thank the anonymous referee for their prompt response and valuable comments which
greatly improved this paper. We also acknowledge productive conversations
within the AGN group at ASTRON.
Funding for the SDSS and SDSS-II has been provided by the Alfred
P. Sloan Foundation, the Participating Institutions, the National
Science Foundation, the U.S. Department of Energy, the National
Aeronautics and Space Administration, the Japanese Monbukagakusho, the
Max Planck Society, and the Higher Education Funding Council for
England. The SDSS Web Site is \url{http://www.sdss.org/}.









\bsp	
\label{lastpage}
\end{document}